\documentclass{aa}  

\usepackage{graphicx}
\usepackage{txfonts}
\usepackage{amsmath}
\usepackage{color}
\usepackage{epsfig}
\usepackage{mathrsfs}
\usepackage{ulem}
\usepackage{balance} 
\usepackage{amssymb,amsmath}
\usepackage{longtable,lscape}
\usepackage{url}
\usepackage{bm}
\DeclareMathAlphabet{\mathbi}{OT1}{ptm}{bx}{it}
\SetMathAlphabet\mathbi{bold}{OT1}{ptm}{bx}{it}

\def\bhm{M_{\bullet}}

\def\calE{{\cal E}}
\def\calI{{\cal I}}
\def\calL{{\cal L}}
\def\calM{{\cal M}}
\def\civ{C {\sc iv}}

\def\cms{\rm cm\,s^{-1}}

\def\mathdotM{\dot{\mathscr{M}}}

\def\qq{q_{_{0.1}}}

\def\Rd{R_{\rm d}}
\def\Rout{R_{\rm out}}

\def\sunm{M_\odot}

\def\mgii{Mg~{\sc ii}}

\def\Rg{R_{\rm g}}

\usepackage{natbib}
\bibpunct{(}{)}{;}{a}{}{,} 

\begin{document} 

\title{Changing-look active galactic nuclei: close binaries of supermassive black holes in action}

\author{Jian-Min Wang\inst{1,2,3} and Edi Bon\inst{4}}

\institute{Key Laboratory for Particle Astrophysics, Institute of High Energy Physics,
Chinese Academy of Sciences, 19B Yuquan Road, Beijing 100049, China, 
\email{wangjm@ihep.ac.cn}\\ \vglue -0.3cm
\and
School of Astronomy and Space Sciences, University of Chinese Academy of Sciences, 19A Yuquan
Road, Beijing 100049, China \\ \vglue -0.3cm
\and
National Astronomical Observatories of China, Chinese Academy of Sciences, 20A Datun Road, 
Beijing 100020, China \\ \vglue -0.3cm
\and Astronomical Observatory Belgrade, Volgina 7, 11060 Belgrade, Serbia
}

\date{Submitted September 8, 2020 / Accepted October 12, 2020}
 
\abstract{
Changing-look active galactic nuclei (CL-AGNs) as a new subpopulation challenge some 
fundamental physics of AGNs because the timescales of the phenomenon can hardly be 
reconciled with accretion disk models. In this Letter{\textit{}}, we demonstrate the 
extreme case: close binaries of supermassive black holes (CB-SMBHs) with high eccentricities 
are able to trigger the CL transition through one orbit. In this scenario, binary black 
holes build up their own mini-disks by peeling gas off the inner edges of the circumbinary 
disk during the apastron phase, after which they tidally interact with the disks during 
the periastron phase to efficiently exchange angular momentum within one orbital period. 
For mini-disks rotating retrograde to the orbit, the tidal torque rapidly squeezes 
the tidal parts of the mini-disks into a much smaller radius, which rapidly results in 
higher accretion and short flares before the disks decline into type-2 AGNs. 
Prograde-rotation mini-disks gain angular momentum from the binary and rotate outward, 
which causes a rapid turn-off from type-1 to type-2. Turn-on occurs around the apastron phase. 
CB-SMBHs control cycle transitions between type-1 and type-2 with orbital periods but allow 
diverse properties in CL-AGN light curves.}

\maketitle
%

\section{Introduction}
Seyfert galaxies are commonly classified into 
type 1 and 2 active galactic nuclei (AGNs) corresponding to ones with and without broad emission 
lines, respectively. The two kinds of AGNs have been successfully unified by obscurations of 
a dusty torus differently orientated to observers \citep{Antonucci1993,Netzer2015}. Despite 
its success, five outliers were detected between the 1970-1980s:  NGC 3516 \citep{Andrillat1968}, 
NGC 7603 \citep{Tohline1976}, Fairall 9 \citep{Kollatschny1985}, NGC 4151 and 3C 390.3 
\citep{Penston1984}, and Mrk 1080 \citep{Cohen1986}. They undergo transitions
from type 1 to type 2 on a timescale of less than one year according to H$\beta$ line, or vice 
versa, they also show \mgii\, 
\citep{Guo2019} and \civ\, \citep{Ross2019} lines.. They are known today as changing-look (CL) 
AGNs (CL-AGNs), which are a fast-increasing population discovered in time domaine surveys 
(e.g., \citealt{MacLeod2016,Yang2018,MacLeod2019,Graham2019,Wang2019,Frederick2019}). 
Their continuum changes from a 
factor of a few to several tens when a type transition occurs. Typical Eddington 
ratios are $\lambda_{\rm Edd}\approx 0.1$ for CL-AGNs in the type 1 state \citep{MacLeod2019}. 
Moreover, CL cannot be explained by obscuration variations according to polarization 
observations \citep{Hutsemekers2019,Marin2019}. This greatly enhances it
as a puzzle. 

Several hypotheses have been suggested to explain CL. Inspired by changes in the 
accretion states of X-ray binaries, \cite{Noda2018} showed that evaporation of the inner part
of the disk drives a change in accretion states into a drop, which causes the CL of Mrk 1018. 
This appears to be supported by the analogy of CL-AGNs to 
X-ray binaries \citep{Ruan2019,Ai2019,Liu2019}. Cooling fronts driven by a sudden change in 
the magnetic torque at the innermost stable circular orbit \citep{Ross2018}, 
or magnetic fields supporting a geometrically thick disk 
\citep{Dexter2019}, or the radiation pressure instability \citep{Sniegowska2020}
lead to a changing state. 
With the standard model of a single SMBH system 
for AGNs \citep{Rees1984}, we can hardly reconcile
the shorter timescales of CL properties with the various instabilities 
of accretion disks because accretion-disk timescales are far longer
 \citep{Stern2018,Ross2018,Lawrence2018}, even if they could last a few 
months, as in the CL-AGN 1ES 1927+654 \citep{Trakhtenbrot2019a}. These facts 
strongly imply an external torque required for 
the fast type transition. Considering  
highly asymmetric profiles of CL-AGNs, we could think that the most natural sources 
driving the transitions are the secondary black hole at the galactic center.

In this Letter, we show one possibility: CB-SMBHs with highly eccentric orbits
can trigger type transitions. In this context,  
the tidal torque of black holes 
acting on the mini-disks can efficiently exchange angular momentum (AM) with the binary 
orbit, giving rise to fast and violent variations in ionizing sources. Consequently, the 
appearance of the AGNs changes. 

\section{Accretion onto binary black holes}
\subsection{Close binaries of SMBHs}
Numerous studies 
have been performed for CB-SMBHs based 
on \cite{Artymowicz1996} and \cite{Armitage2002}. The basic conclusions 
are that 1) the circumbinary disk (CBD) surrounding the CB-SMBHs has a central cavity  
with a radius of 
$R_{\rm CBD}\approx 2a(1+e)$, which is formed by tidal torque of the binary with 
a half-major axis ($a$) and eccentricity ($e$); 2) each black hole efficiently peels 
streams of gas off the inner edge of the CBD that rapidly traverse the
cavity \citep{MacFadyen2008,Roedig2014,Farris2015a,Farris2015b,Shi2015,Bowen2018}; and
3) the peeled streams form two mini-disks around each member of the binary 
\citep{Roedig2014,Farris2014,Farris2015a,Farris2015b,DOrazio2016,Tang2018,dAscoli2018,Moody2019,
Bowen2018,Bowen2019,Munoz2020}. 

We consider one CB-SMBH system,
whose separation with phases is described by $D=a(1-e^{2})/\left(1+e\cos\theta\right),$ and
$\theta$ is the orbital phase angles. Periastron and apastron are given by
$D_{\rm min, max}=a(1\mp e)$, respectively, where tidal interaction is strongest and weakest. 
The CB-SMBH periods are 
%
\begin{equation}\label{eq:Porb}
P_{\rm orb}=\left\{\begin{array}{l}\vspace{1ex}
19.5\,a_{\rm 3.4k}^{3/2}M_{8}(1+q)^{-1/2}\,{\rm yr},\\
3.1\,a_{\rm 1k}^{3/2}M_{8}(1+q)^{-1/2}\,{\rm yr},
\end{array}\right.
\end{equation}
where $a_{\rm 1k,3.4k}=a/(1,3.4) \times  10^{3}\Rg$ was chosen for two types of eccentricities
to create a turn-off (for a few years of viscosity time for the innermost part of the disks, 
see Eq.\ref{eq:Rtid} and \ref{eq:tvis}, and also for high-$e$ orbits), 
$\Rg=GM_{\bullet}^{p}/c^{2}=1.5\times 10^{13}\,M_8\,$cm is the gravity radius, 
$M_{8}=M_{\bullet}^{p}/10^{8}\sunm$ is
the primary black hole mass, $q=M_{\bullet}^{s}/M_{\bullet}^{p}$ is the mass ratio, 
$G$ is the gravitational constant, and $c$ is the speed of light. H$\beta$-BLR (broad-line region)
radius follows $R_{\rm H\beta}/\Rg\approx 
6.4\times 10^{3}\left(\eta_{0.1}/\kappa_{10}\right)^{1/2}(\mathdotM/M_{8})^{-1/2}$
\citep{Bentz2013,Du2019}, indicating that the BLR is located in the CBD, where 
$\eta_{0.1}=\eta/0.1$ is the radiation efficiency and $\kappa_{10}=\kappa_{\rm bol}/10$ is the 
bolometric factor,
$\mathdotM=\dot{M}_{\bullet}c^{2}/L_{\rm Edd}$ is the dimensionless accretion 
rate ($\dot{M}_{\bullet}$), and $L_{\rm Edd}$ is the Eddington luminosity. 
CB-SMBHs are assumed to have two solitary BLRs
\cite[e.g.,][]{Shen2010,Popovic2012,Wang2018}.

\subsection{Building-up of mini-disks}
As characteristic configurations, we show the simplest four coplanar cases of mini-disks 
and orbital AM of CB-SMBHs in Fig. 1. The complexity arises from the random accretion 
of peeled gas and supply onto the SMBHs. There are solid reasons for fully chaotic 
and random accretion onto CB-SMBHs making both prograde and retrograde accretions 
possible in the context of merged galaxies (\citealt{Nixon2011,Nixon2013,Dunhill2014}, 
see also for details in \citealt{Roedig2012,Roedig2014} and hereafter). Observationally, 
there is increasing evidence for random accretion. SINFONI observations, for example, 
showed two tongues from the south and north in $\sim 10\,$pc 
regions in NGC\,1068 \citep{Muller2009}, dual AGNs with an antiparallel AM 
in Arp 220 \citep{Wheeler2020}, and other evidence \citep{Tremblay2016,Temi2018}.
The inner edge of the CBD significantly differs from the accretion disk around 
a single black hole: 1) part of the lumps (from the $m=1$ mode) receive AM from 
the binary and are flung at the inner edge; but 2) via the turbulence of the CBD disk, 
AM is transferred outward, which drives the lumps inward. Moreover, the peeled gas has 
a very low AM on average, but the AM direction of individual clumps might be random to 
some degree.

Following \cite{Roedig2014b}, we assumed that the pealing accretion rates of the primary and secondary 
black holes are $\dot{M}_{p}=f_{p}\dot{M}_{\bullet}$ and $\dot{M}_{s}=f_{s}\dot{M}_{\bullet}$, 
respectively, but $f_{p}+f_{s}\lesssim 1$, where $\dot{M}_{\bullet}$ is the accretion rate of the CBD,
$f_{p}$ and $f_{s}$ are fractions depending 
on the orbits of the CB-SMBHs. For simplicity, we assumed them to be free parameters.
We roughly divided the orbit into two parts: 1) 
the apastron phase ($f_{t}P_{\rm orb}; f_{t}\approx 1$), and 2) the periastron phase 
of $t_{\rm peri}\approx (1-e)^{3/2}P_{\rm orb}\lesssim 0.03 P_{\rm orb}$ 
for $e\gtrsim 0.9$. The mini-disk is accumulating to have a mass of 
$\Delta M_{p}=f_{t}P_{\rm orb}f_{p}\dot{M}_{\bullet}$ during the apastron phase (tide 
is weak), but the disk strongly interacts with individual black 
holes during the periastron phase (the peeling is negligible). It has been shown that
the peeling rates are significantly modulated by eccentric orbits \citep{Bogdanovic2008}, 
but only the total peeled mass is important for exploring whether
CB-SMBHs can make a type transition due to tidal interaction. 

\begin{figure*}\label{fig:AM}
   \centering
   \vglue -0.3cm
   \includegraphics[angle=0,width=0.859\textwidth,trim = -50 -20 10 0, clip]{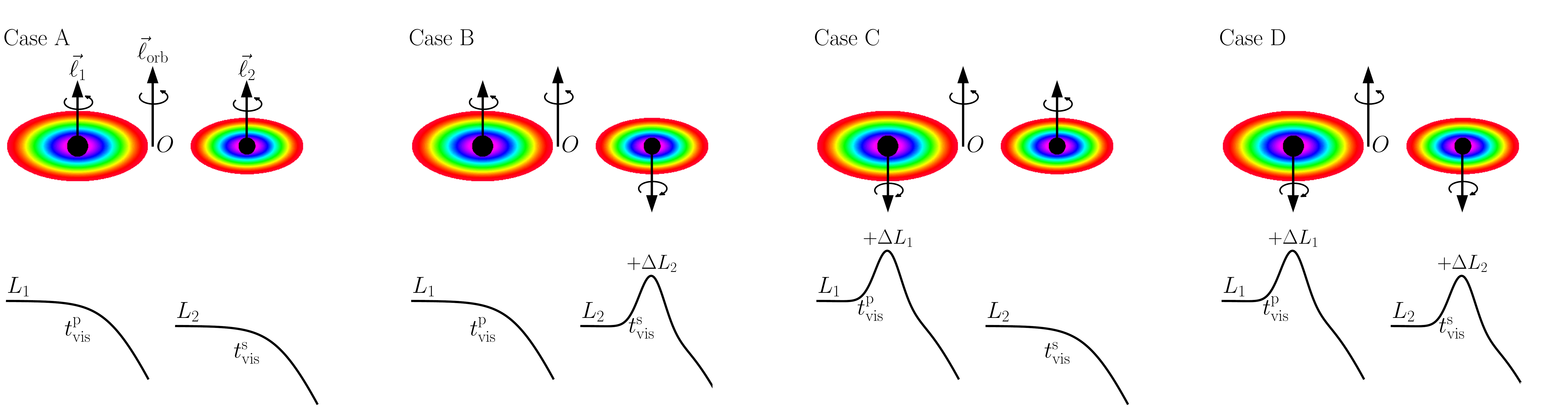}
   \vglue -0.35cm
   \caption{\footnotesize
   Simplest AM configurations (AMC) of CB-SMBHs and corresponding light curves when 
   the companions pass through periastron. 
   $\vec{\ell}_{\rm p},\vec{\ell}_{\rm orb}, \text{and } \vec{\ell}_{\rm s}$ denote the AM 
   of the primary, the orbital, and the secondary black holes, respectively. $O$ is the mass 
   center of the binary.
   Tidal interaction between mini-disks and SMBHs leads to two types of quenching modes: 
   1) fast gains of AM for prograde mini-disks (Case A) from the binary orbit, and  
   2) precursor-flaring and quenching for a retrograde mini-disk (the secondary in Case B).
   The precursor flaring is caused by the rapid loss of AM of the
   tidal part so that accretion in the remainder of the mini-disk is accelerated. Here
   $\Delta L_{1,2}=\eta \left(\delta M_{\rm t}^{\rm p,s}/t_{\rm v}^{\rm p,s}\right)c^{2}$ 
   represents the increased luminosity 
   due to the enhanced accretion for the primary and the secondary BHs, respectively (see 
   text for details).
   }
   \vglue-0.3cm
\end{figure*}

The peeled gas forms mini-disks around each black hole. 
During one apastron phase, the peeled gas is
\begin{equation}\label{eq:mass_peel}
\frac{\Delta M_{\rm peel}}{\sunm}=\left\{\begin{array}{l}
4.4\,M_{8}f_{p}\mathdotM f_{t}P_{20}\\ 
0.4\,\qq M_{8}f_{s}\mathdotM f_{t}P_{20}
\end{array}\right.
\end{equation}
for the primary and the secondary, respectively, where $q_{0.1}=q/0.1$ and 
$P_{20}=P_{\rm orb}/20{\rm yr}$. 
At the inner edge, the cooling timescale 
of the peeled gas is $t_{\rm cool}\approx 5.6\,T_{4}^{1/2}n_{10}^{-1}\,$hours, where 
$T_{4}=T/10^{4}$K, and $n_{10}=n/10^{10}{\rm cm^{-3}}$ is the density. We find that 
$t_{\rm cool}$ is much shorter than the free fall of 
$t_{\rm ff}\approx 4.0\,r_{\rm 4k}^{3/2}M_{8}\,$yr $\lesssim P_{\rm orb}$,
where $r_{\rm 4k}=R/4\times 10^{3}\Rg$. Neglecting the complicated processes, 
we assumed the mini-disks to be described by the inner region of the standard 
model \citep{Shakura1973}. The surface density, mid-plane temperature,
and radial velocity of the disks are \citep[see][]{Kato1998}
\begin{equation}\label{eq:sigma}
\left\{
\begin{array}{l}
\Sigma=3.5\times 10^{2}\,\alpha_{0.1}^{-1}\mathdotM^{-1}r^{3/2}\,{\rm g\,cm^{-2}},\\ 
T_{c}=1.1\times10^{6}\,\left(\alpha_{0.1}M_{8}\right)^{-1/4}r^{-3/8}\,{\rm K},\\
\upsilon_{r} =4.3\times 10^{8}\,\alpha_{0.1}\mathdotM^{2}r^{-5/2}\,{\rm cm\,s^{-1}},
\end{array}\right.
\end{equation}
provided $\mathdotM\ge 10^{-2}(\alpha_{0.1}M_8)^{-1/8}$, and this works within the outer radius of 
$1.7\times 10^{2}\left(\alpha_{0.1}M_{8}\right)^{2/21}\mathdotM^{16/21}$, 
where $\alpha_{0.1}=\alpha/0.1$ is the viscosity.  The viscosity timescale (or radial motion one) 
is
\begin{equation}\label{eq:tvis}
t_{\rm vis}\approx 3.5\,\alpha_{0.1}^{-1}M_{8}\mathdotM^{-2}r_{10}^{7/2}\,{\rm yr},
\end{equation}
where $r_{10}=R/10\Rg$, which is sensitive to $\mathdotM$ and the radius. Given $r=r_{*}=10$, 
we have $t_{\rm vis}=3.5\,M_{8}\,{\rm yr,}$ denoted $t_{\rm vis}^{r_*}$ for $\alpha=0.1$
and $\mathdotM=1$, which agrees with the characterized transition of CL-AGNs. 
In order to create a CL, the suggested external torque has to
remove the AM of the $R\gtrsim 10\Rg$ disk.
 
By integrating the surface density of Eq. (\ref{eq:sigma}), we obtained a mini-disk mass of 
$\Delta M_{\rm disk}/\sunm\approx2.5\,\alpha_{0.1}^{-1}\mathdotM^{-1}M_{8}^{2}r_{20}^{7/2}$, 
where $r_{20}=R/20\Rg$, yielding outer boundary radii of the primary and the secondary mini-disks
from $\Delta M_{\rm disk}=\Delta M_{\rm peel}$,
\begin{equation}\label{eq:Rout}
\frac{\Rout}{\Rg}\approx\left\{\begin{array}{ll}
23.5\,\alpha_{0.1}^{2/7}\mathdotM^{4/7}M_{8}^{-2/7}P_{20}^{2/7}f_{p}^{2/7}f_{t}^{2/7},\\
45.3\,\alpha_{0.1}^{2/7}\mathdotM^{4/7}(\qq M_{8})^{-2/7}P_{20}^{2/7}f_{s}^{2/7}f_{t}^{2/7},
\end{array}\right.
\end{equation}
respectively. Circularization of the peeled gas relies on its AM, but 
our model takes a conservative estimate of low AM (because $R_{\rm out}\ll R_{\rm CBD}$).
The disks retain vertical equilibrium with a very short timescale of 
$t_{\rm eq}
=\Omega_{\rm K}^{-1}=0.5\,M_{8}r_{20}^{3/2}\,{\rm days}$, which is
equal to the free-fall timescale ($t_{\rm ff}$) after the accretion disk AM is 
removed. With the sound speed of $c_{s}\approx 10^{4}\,T_{c}^{1/2}\cms$, 
we have a radial dynamical timescale of
$t_{\rm dyn}
            =  1.6\,\alpha_{0.1}^{1/8}M_{8}^{9/8}r_{20}^{19/16}\,{\rm yr}$ for
%
keeping a whole disk under tidal perturbations.
Moreover, 
the viscosity time scales of the mini-disks are 
$t_{\rm vis}\approx (69.6,692.5)\,\alpha_{0.1}^{-1}M_{8}\mathdotM^{-2}\,$yr at 
$R_{\rm out}$ given by Eq.(\ref{eq:Rout}), which are much longer than $P_{\rm orb}$. 
The peeled-gas rates are modulated 
by orbital motion \citep[e.g.,][]{Bogdanovic2008,Miranda2017} with about $P_{\rm orb}$,
but the modulated rates are smeared by the viscosity so that temporal radiation (light 
curves) from the mini-disks does not follow the orbital modulations.
Thus only the total peeled mass is important for the current issues. 

\subsection{Tidal torque from a black hole}
The effects of tidal torques acting on mini-disks depend on the AMC. 
For a prograde mini-disk ($\vec{\ell}_{i}\!\parallel\!\vec{\ell}_{\rm orb}$, $i=1,2$), the tidal 
torque will physically divide the mini-disk into two parts at radius $R_{\rm tid}$ 
(in Eq. \ref{eq:Rtid}). 
In such a case, the inner part of the mini-disks ($\le R_{\rm tid}$) is tidal free, whereas 
its tidal part is accelerated through delivering orbital AM to the part, and it 
becomes a ``decretion disk'', leading to dramatical decrease in accretion rates 
with $t_{\rm vis}^{r_*}$ and quenching accretion. For a retrograde mini-disk, the tidal 
part in contrast efficiently looses AM and
is rapidly squeezed into the tidal radius ($\sim t_{\rm ff}$), dramatically accelerating 
the ongoing accretion rates and creating a flare until the inner part diminishes. The tidal
interaction between the black holes and the mini-disks causes diverse phenomena
of CL-AGNs. 

The tidal force acting on the secondary disk by the primary black hole is 
$F_{\rm tid}\approx 
\pm G\bhm^{p} \delta M_{\rm tid}{D^{-2}}\left(R_{\rm d}/{D}\right)$,
where $\delta M_{\rm tid}$ is the mass of the tidal part at radius $\Rd$. 
This net tidal force depends on the AMCs in Fig. 1. 
Here $\text{the plus-minus sign}$ represents gain and loss of AM from 
tidal interaction for prograde and retrograde mini-disks, respectively. 
The tidal torque is 
\begin{equation}\label{eq:torque}
{\cal T}_{\rm tid}\approx \pm\delta M_{\rm tid}c^{2}\left(\frac{\Rg}{D}\right)\left(\frac{\Rd}{D}\right)^{2}.
\end{equation}
The AM of the tidal part is approximated by
%
$\delta{\cal L}_{\rm d}\approx \delta M_{\rm tid}\ell_{\rm d}$, where 
$\ell_{\rm d}=\sqrt{G\bhm^{s} R_{\rm d}}$ is the specific AM,
%
yielding the tidal radius from the balance given by 
$\delta {\cal L}_{\rm d}={\cal T}_{\rm tid}t_{\rm peri}$
\begin{equation}\label{eq:Rtid}
\frac{R_{\rm tid}}{\Rg}=\left(\frac{\xi_{q}}{4\pi^{2}}\right)^{1/3}
                        \left(\frac{a{\cal E}}{\Rg}\right)
           =\left\{\begin{array}{l}\vspace{1ex} 
           14.4\,\xi_{0.11}^{1/3}a_{\rm 3.4k}\calE_{0.03},\\
           14.1\,\xi_{0.11}^{1/3}a_{\rm 1k}\calE_{0.1},
           \end{array}\right.           
\end{equation}
where $\xi_{q}=q(1+q)$, $\xi_{0.11}=\xi_{q}/0.11$ for $q=0.1$, $\calE=(1-e)$ and 
$\calE_{0.03,0.1}=\calE/(0.03,0.1)$ for $e=(0.97,0.9)$, respectively.

Density waves excited by the torque in the tidal part
are strong enough to form shocks and to transport AM \citep{Spruit1987}. Recent
simulations showed that this is a powerful way of transporting AM outward \citep{Ju2016},
even stronger than the known. 
With the help of $\delta \calL_{\rm d}$, we have the timescale within which the AM is exchanged, 
$t_{\rm tid}=\delta \calL_{\rm d}/\left|{\cal T}_{\rm tid}\right|$,
\begin{equation}\label{eq:ttid}
t_{\rm tid}=q^{1/2}\frac{\Rg}{c}\left(\frac{D}{\Rg}\right)^{3/2}\left(\frac{\Rd}{D}\right)^{-3/2},
\end{equation}
with the wave density pattern velocity $V_{\rm sp}$. 
The mini-disks respond to the tidal perturbations with $t_{\rm dyn}$. The 
spiral shocks excited by the tidal torque are nonlocal with Mach numbers of 
$\calM=V_{\rm sp}/c_{s}=t_{\rm dyn}/t_{\rm tid}$ beyond a radius at $\calM=1,$
\begin{equation}\label{eq:Rsh}
\frac{R_{\rm sh}}{\Rg}=\left\{\begin{array}{l}\vspace{1ex}
5.9\,\alpha_{0.1}^{-2/43}\qq^{8/43}M_{8}^{-2/43}a_{\rm 3.4k}^{48/43}\calE_{0.03}^{48/43},\\
5.8\,\alpha_{0.1}^{-2/43}\qq^{8/43}M_{8}^{-2/43}a_{\rm 1k}^{48/43}\calE_{0.1}^{48/43},
\end{array}\right.
\end{equation}
and the propagation timescale ($t_{\rm sh}=t_{\rm dyn}=t_{\rm tid}$) is
\begin{equation}\label{eq:ttid}
t_{\rm sh}=\left\{\begin{array}{l}\vspace{1ex}
0.37\,\alpha_{0.1}^{3/43}\qq^{19/86}M_{8}^{46/43}a_{\rm 3.4k}^{57/43}\calE_{0.03}^{57/43}\,{\rm yr},\\ \vspace{1ex}
0.34\,\alpha_{0.1}^{3/43}\qq^{19/86}M_{8}^{46/43}a_{\rm 1k}^{57/43}\calE_{0.1}^{57/43}\,{\rm yr}.
\end{array}\right.
\end{equation}
Here we take $D=D_{\rm min}$ for the shortest timescale. We find that $t_{\rm sh}< t_{\rm vis}$
, indicating that spiral shocks can efficiently exchange the AM of the mini-disk with the binary. 
Moreover, the condition of $t_{\rm vis}\lesssim P_{\rm orb}$ is necessary for a turn-off of 
accretion onto SMBHs. Combining Eqs.(\ref{eq:Porb},\ref{eq:tvis},\ref{eq:Rtid}), we have
\begin{equation}\label{eq:mathdotM}
\mathdotM \gtrsim \left\{\begin{array}{l}\vspace{1ex}
0.8\,\alpha_{0.1}^{-1/2}\calE_{0.03}^{7/4}q_{0.1}^{7/12}(1+q)_{1.1}^{5/6}a_{\rm 3.4k},\\ \vspace{1ex}
1.9\,\alpha_{0.1}^{-1/2}\calE_{0.1}^{7/4}q_{0.1}^{7/12}(1+q)_{1.1}^{5/6}a_{\rm 1k},
\end{array}\right.
\end{equation}
implying $\lambda_{\rm Edd}=\eta\mathdotM\gtrsim0.1$,
otherwise the turn-off cannot be achieved within one orbital period. Considering the large 
uncertainties of $\bhm$ and the bolometric luminosity, Eq.(\ref{eq:mathdotM})
is roughly consistent with the $\lambda_{\rm Edd}\gtrsim0.1$ CL-AGNs 
\citep[see Fig. 6 in][]{MacLeod2019}, showing the validity of this turn-off mechanism.

For the prograde case, on the other hand, the decretion disk could be changed into elliptical 
shapes so that some of it will be accreted onto SMBHs, leading to an inefficient turn-off. We keep
this effect in mind for numerical simulations in the future.

\subsection{Gravitational waves}
Owing to gravitation waves (GWs), the binary orbit shrinks on a timescale given by
\citep[see Eq. 5.6 in][]{Peters1964}, 
\begin{equation}
t_{\rm GW}\approx\left\{\begin{array}{l}\vspace{1ex}
   1.8\times 10^{4}\,\xi_{0.11}^{-1}M_{8}a_{\rm 3.4k}^{4}\calE_{0.03}^{7/2}\calI_{4.2}^{-1}\,{\rm yr},\\ \vspace{1ex}
   8.9\times 10^{3}\,\xi_{0.11}^{-1}M_{8}a_{\rm 1k}^{4}\calE_{0.1}^{7/2}\calI_{3.7}^{-1}\,{\rm yr},
            \end{array}\right.
\end{equation}
corresponding to $N_{\rm orb}\approx(908,2881)$ orbits, where
$\calI_{e}=1+73e^{2}/24+37e^{4}/96$, and $\calI_{4.2,3.7}=\calI_{e}/(4.2,3.7)\,$ for $e$ 
previously given. The intrinsic strain amplitude is
\begin{equation}
h_{s} =7.0\times 10^{-18}\,\qq(1+q)^{-1/3}M_{8}^{5/3}P_{20}^{-2/3}d_{100}^{-1},
\end{equation}
from $h_{s}=(128/5)^{1/2}(GM_{c})^{5/3}(\pi f_{\rm GW})^{2/3}/c^{4}d_{\rm L}$, 
$f_{\rm GW}=2/P_{\rm orb}\approx 32\,P_{20}^{-1}\,$nHz 
is the frequency of GWs, $M_{c}=q^{3/5}{\bhm^{p}}(1+q)^{-1/5}$ is the chirping mass,\
and $d_{100}=d_{\rm L}/100\,{\rm Mpc}$ is the distance \citep[e.g.,][]{Graham2015}. 
For pure GW-driven evolution, $t_{e}\approx (32.7,9.3)t_{\rm GW}$ , and the initial 
$e_{0}=(0.97,0.9)$ will reduce to $(0.95,0.84)$ by dropping half of $a$, where $t_{e}=\left(d\ln e/dt\right)^{-1}$ is the evolutionary timescale of eccentricity.
However, a monotonic $e$-gain (for $e\gtrsim 0.6$) with accretion of the secondary black hole 
in a retrograde binary can keep high-$e$ orbits prior to the merger \citep{Amaro-Seoane2016}.
With the complicated $e$-evolution, the high-$e$ orbits can
be kept in some orbits that result in CL. 

We focus on the extreme cases here in which CL transition is triggered by
one tidal interaction during one orbit. If the initial $e$ of CB-SMBH orbits 
is not as high as $e\approx 0.9$, the binaries may undergo tidal interaction for several 
orbits and lead to CL transition, or they undergo increases in $e$ in the case of retrograde
accretion, which finally leads to a transition. In this context, the mini-disks can
accumulate more during several orbital periods; CL-AGNs have different
variations from what we discussed here. A future paper will study  
the intermediate-$e$ cases.

\section{Turn-on and turn-off}
 As shown
in Fig. 1,  we focused on four simplest configurations: 
(1) Case-A ($\uparrow,\uparrow,\uparrow$); (2) Case-B ($\uparrow,\downarrow,\uparrow$);
(3) Case-C ($\uparrow,\downarrow,\downarrow$), and (4) Case-D ($\downarrow,\downarrow,\uparrow$).
We neglected spins of black holes, avoiding more configurations.  
We note that the appearance of Cases B and C could be less common
than that of Cases A and  D.
For a convenient discussions, we use $\delta M_{t}^{p,s}$ as 
the masses of the tidal parts of the primary and secondary mini-disks, and 
$\delta M_{0}^{p,s}$ as the mass of the remaining parts. We stress that the discussions are only valid 
for the characterized properties of the process.

\subsection{Case A}
In Case A, both mini-disks are prograde. Tidal torque transfers orbital AM
to both disks simultaneously so that the tidal parts 
(with mass of $\delta M_{t}^{p,s}$) become
decretion disks that rotate outward \citep[e.g.,][]{Rafikov2016}, and accretion onto
the binary is 
interrupted after the mass ($\delta M_{0}^{p,s}$) of the nontidal part 
is depleted. 
We call this the rest accretion, which radiates at luminosities of 
$\Delta L_{\rm A}=\eta (\delta M_{0}^{p}/t_{\rm vis}^{p}+\delta M_{0}^{s}/t_{\rm vis}^{s})c^{2}$,
where $t_{\rm vis}^{p,s}$ are viscosity timescales of the primary and secondary mini-disks.
According to the standard disk model, photons ($\epsilon$) are emitted from  
$R_{\epsilon}/\Rg\approx 6\,(\dot{\mathscr{M}}/M_{8})^{1/3}(\epsilon/13.6{\rm eV})^{-4/3}$
as the innermost regions of 
the mini-disks. The expansion of the tidal part can simply be estimated by 
$\Delta R_{\rm d}/R_{\rm d}\approx 2\Delta(\delta\calL_{\rm d})/\delta\calL_{\rm d}=2$
because the balance of $\delta {\cal L}_{\rm d}={\cal T}_{\rm tid}t_{\rm peri}$.
As a consequence, gas supply to the $R_{\epsilon}$-region
is interrupted by the torque, leading to violent variations in both UV and H$\beta$ line.
Turn-off occurs then with a transition from type 1 to  type 2 after periastron. 

Turn-on transition starts from accumulation via peeling 
gas from the CBD after the periastron phase. First, the tidal part begins to 
fall back, but on a timescale that is longer than $t_{\rm vis}$ by a factor of
$(\Delta R_{\rm d}/R_{\rm d})^{7/2}\approx 11.3$ ($>P_{\rm orb}$). Second, peeled gas
continues to accumulate because the outward-rotating gas is still bounded by its black hole 
in the same disk plane (because $t_{\rm eq}\ll t_{\rm dyn}$). This results in continuous 
growth of the mini-disks from peeled gas over every orbiting cycle until the mini-disk 
radii exceed the Roche radius given by $R_{\rm H}\approx 0.27q^{\mp0.3}a$ for the primary 
and secondary mini-disks \citep{Bowen2017}. With limitation of GW-driven
timescales of orbit decays, on the other hand, the CB-SMBHs will undergo cycles of 
$n_{\rm cyc}\approx t_{\rm GW}/P_{\rm orb}\sim 10^{3}$ around
the apastron phase for $P_{\rm orb}\sim 20\,$yr and $t_{\rm GW}\sim 2\times 10^{4}\,$yr. 
Mini-disks then grow to an outer radius of $R_{\rm out}/\Rg\approx(88,170)$
(by multiplying the factor $f_{t}$ with $n_{\rm cyc}$ in Eq.\ref{eq:Rout})
but $R_{\rm out}/\Rg\le R_{\rm H}$ for the primary and secondary.
This disk is able to produce optical continuum like normal AGNs. 
The innermost regions undergo violent variations that cause
fast and large variabilities only in the UV, but they are relatively less variable 
in the optical, agreeing with that of CL-AGNs 
\citep{Ross2018,Lawrence2018}.

The growth of prograde mini-disks results from the circularization of the peeled gas 
through AM composition. 
Kinetic energy isdissipated in this process to radiate some energy 
(but the spectral energy distributions depend on the composition). Consequently, 
this causes complicated variations in the continuum on various timescales. More 
work needs to be done for details through numerical simulations in the future.

\subsection{Cases B and C}
In Case B, the primary mini-disk is prograde to the orbit, but the secondary is retrograde. 
The primary disk receives AM from the orbit, and the UV emission
declines as in Case A, but the secondary shows a flare in UV before 
quenching. This is caused by a fast increase in the accretion rates within $R_{\rm tid}$ 
through a rapid squeezing of the $\delta M_{\rm t}^{s}$ caused by loss of its AM. 
The enhanced accretion rates can be roughly estimated by 
%
$\dot{M}_{s}\approx 
                    \left(\delta M_{0}^{s}+\delta M_{t}^{s}\right)/{t_{\rm vis}^{s}}$.
%
The enhanced accretion leads to an increased luminosity 
$\Delta L_{2}=\eta \left(\delta M_{\rm t}^{s}/t_{\rm vis}^{s}\right)c^{2}$. 
After fast accretion within $R_{\rm tid}$, the mini-disk is 
quenched, and consequently, turn-off occurs on a timescale of $\sim 1$yr (Eq. \ref{eq:tvis}). 
Interestingly, 
 a flare like this was found as a precursor of CB-SMBHs in numerical simulations by \cite{Chang2010}.
In this case, the primary contributes optical emissions, but both components contribute to 
violent variations in UV emissions. In particular, the gas squeezed by the tidal torque undergoes supersonic infall and may lead to $\gamma$-ray emission on short timescales 
\citep{Kolykhalov1979,Meszaros1983}. This might be detectable by {\it Fermi}. 

Case C is similar to Case B, but the primary mini-disk in Case C is retrograde to the orbit. It first
undergoes a flare and then decays. The secondary mini-disk is quenched as in Case A, which 
is driven to rotate outward, but it causes optical emissions similar to Case B. 
We note that the flaring and quenching timescales of the two mini-disks
might be different depending on black hole masses, accretion rates, and the binary orbit.

\subsection{Case D}
In Case D, both mini-disks are retrograde. Both tidal parts of the disks 
are squeezed rapidly into $\lesssim R_{\rm tid}$ regions, which accelerates the accretion rates 
of the disks. Then a turn-off transition occurs. In this case, the total 
accretion rates of the flare luminosity of the binary are about
%
$\dot{M}_{\rm tot}=\dot{M}_{p}+\dot{M}_{s}$,
%
where $\dot{M}_{p,s}=\left(\delta M_{0}^{p,s}+\delta M_{t}^{p,s}\right)/t_{\rm vis}^{p,s}$.
The enhanced accretion depends on
the black hole mass, and thus the two flares of the mini-disks have different timescales.
These characterized light curves depend on the ratios of 
$\delta M_{\rm t}^{p,s}/\delta M_0^{p,s}$ and their relative viscosity timescales.
For cases of extremely eccentric orbits, the mini-disks may be completely tidal destroyed 
if $R_{\rm tid}\sim 6 \Rg$ for extreme cases with very high$-e$ orbits. In this case, 
the flares are stronger because
all mini-disks are squeezed into the black hole.

Real situations of mini-disks or orbits in CB-SMBHs may be more complicated 
than that in Fig. \ref{fig:AM}. They might be warped and noncoplanar, such as 
in OJ 287 \citep{Sillanpaa1988}. 
Moreover, the relative appearance of the four cases remains open, 
but Cases A and D might be more common than others. 
Finally, we would pointed out that the current scenario provides a cycle 
of CL transition with orbit period in the nonperiodical light curves of
the continuum. Turn-on and turn-off timescales are determined 
by the orbit, but they might have a quasi-periodicity when 
the CBD is self-gravitating to form clumps.

\section{Conclusions and discussions}
The rapidity of the change in CL-AGNs is a puzzle. Based on heuristic information 
from observations, we demonstrated the extreme case in which CB-SMBHs with low mass ratios
and high eccentricities might cause a fast transition of states within one orbit. Both 
black holes peel gas off the inner edge of the circumbinary disk to form mini-disks around 
them. In a system like this, the tidal torque on the mini-disks caused by companion black 
holes is strong enough to exchange AM between the disks and the binary orbit
for rapid changes in the innermost regions of the disk. The mini-disks are 
either being squeezed into the innermost regions or are expanded outward (to 
form a decretion disk), depending on the AMC
of the binary system. The different configurations cause the diverse properties 
of CL-AGNs, including flaring precursors of type transitions. CB-SMBHs drive the 
changing look of AGNs to appear in transition cycles, likely with orbital periods (or quasi-periodical 
transition due to complications at the inner edge of the CBD). This scenario can be 
tested by the turn-on and turn-off behaviors in historical data of spectral variations
over several decades. CL-AGNs might contribute a significant fraction of the low-frequency
gravitational wave background.

Several aspects should be investigated to explain the nature of CL-AGNs. First, 
details of their host morphology should be searched for signatures of disrupted or ripples 
as may be caused by merged remnants. This is inspired by results reported by \cite{Charlton2019}, who clearly found such 
features in {\it Gemini} images of four CL-quasars. Second, CL-AGNs could be characterized 
by profile asymmetries in their broad emission lines. We visually confirmed the broadline 
profiles of CL-AGNs and found that all{\it } of them have either highl -asymmetric or 
double-peaked profiles, but the reverse is not supported by current observations.  
AGNs are monitored with the H$\beta$ Asymmetry (MAHA), which is a long-term campaign searching for CB-SMBHs
\citep{Du2018}. We hope to see signals of CB-SMBHs from their 2D transfer 
functions \citep[][see a recent review of Wang \& Li 2020]{Wang2018,Kovacevic2019,Songsheng2020}
as well as signals revealed by 
GRAVITY/VLTI \citep{Songsheng2019,Kovacevic2020}. CL-AGNs could be the best candidates
of CB-SMBHs. Third, the spectral energy distributions 
of CL-AGNs are poorly understood at epochs of type transitions. CB-SMBHs drive 
far more violent variations in the UV than in the optical, which apparently supports our scenario, but 
the complicated processes of the tidal interaction may lead to $\gamma$-rays from the inflows with 
low AM. Monitoring campaigns of CL-AGN multi-$\lambda$ continuum will 
greatly help understand the physics of these processes.

We will study the temporal details of the tidal interaction and accretion rates of the 
mini-disks in a binary orbit in future work. This allows us to compare with observed light 
curves of CL-AGNs or to predict some precursors of CL transition, and reveal the physics of the peeling between the binary and the CBD from observations. In this context,
time-dependent accretion (and decretion) disks should be employed, including the
spectral energy distributions. Finally, CL-AGNs might significantly contribute to
the GW background, and this will be studied in the future.

\begin{acknowledgements}
We are grateful to an anonymous referee, and T. Bogdanovi\'c for helpful discussions as well as
Z. Yu, Y.-R. Li and Y.-Y. Songsheng from the IHEP AGN group.
JMW thanks the support by National Key R\&D Program of China through grant -2016YFA0400701, 
by NSFC through grants NSFC-11991050, -11991054, -11833008, -11690024, and by grant 
No. QYZDJ-SSW-SLH007 and No.XDB23010400.
EB acknowledges the Serbian support 
through contract number 451-03-68/2020-14/200002, and CA16104.
\end{acknowledgements}

\end{document}